\documentclass[twocolumn, showpacs, longbibliography, prx, aps, amsmath, amssymb,superscriptaddress]{revtex4-2}
\usepackage{amsfonts} 
\usepackage{graphicx} 
\usepackage[colorlinks,linkcolor=blue,citecolor=blue,urlcolor=blue]{hyperref}
\usepackage{float}
\usepackage[normalem]{ulem}
\usepackage[switch,columnwise]{lineno}

\begin{document}

\author{Zhoumuyan Geng}
\affiliation {Center for Nanophotonics, AMOLF, Science Park 104, 1098 XG Amsterdam, The Netherlands}

\author{Johanna Theenhaus}
\affiliation {Center for Nanophotonics, AMOLF, Science Park 104, 1098 XG Amsterdam, The Netherlands}

\author{Biplab K. Patra}
\affiliation {Center for Nanophotonics, AMOLF, Science Park 104, 1098 XG Amsterdam, The Netherlands}

\author{Jian-Yao Zheng}
\affiliation {Center for Nanophotonics, AMOLF, Science Park 104, 1098 XG Amsterdam, The Netherlands}

\author{Joris Busink}
\affiliation {Center for Nanophotonics, AMOLF, Science Park 104, 1098 XG Amsterdam, The Netherlands}

\author{Erik C. Garnett}
\affiliation {Center for Nanophotonics, AMOLF, Science Park 104, 1098 XG Amsterdam, The Netherlands}

\author{Said Rahimzadeh Kalaleh Rodriguez}  \email{s.rodriguez@amolf.nl}
\affiliation {Center for Nanophotonics, AMOLF, Science Park 104, 1098 XG Amsterdam, The Netherlands}

\title{Fano lineshapes and Rabi splittings: Can they be artificially generated or obscured by the numerical aperture?}


\begin{abstract}
Fano resonances and Rabi splittings are routinely reported in the scientific literature. Asymmetric resonance lineshapes are usually associated with Fano resonances, and two split peaks in the spectrum are often attributed to a Rabi splitting.  True Fano resonances and Rabi splittings are unequivocal signatures of coherent coupling between subsystems. However, can the same spectral lineshapes characterizing Fano resonances and Rabi splittings arise from a purely incoherent sum of intensities?  Here we answer this question through experiments with a tunable Fabry-P\'erot cavity containing a CsPbBr\textsubscript{3} perovskite crystal. By measuring the transmission and photoluminescence of this system using microscope objectives with different numerical aperture ($NA$), we find that even a modest $NA=0.4$ can artificially generate Fano resonances and Rabi splittings.  We furthermore show that this modest $NA$ can obscure the anti-crossing of a bona fide strongly coupled light-matter system. Through transfer matrix calculations we confirm that these spectral artefacts are due to the incoherent sum of transmitted intensities at different angles captured by the $NA$. Our results are relevant to the wide nanophotonics community, characterizing dispersive optical systems with high numerical aperture microscope objectives. We conclude with general guidelines to avoid pitfalls in the characterization of such optical systems.
\end{abstract}

\maketitle

Fano resonances and Rabi splittings have inspired countless efforts in photonics research~\cite{Miroshnichenko10, Nordlander10, Torma14, Rybin17}. These two effects were discovered in quantum frameworks, yet their essence can be easily recognized in classical models of coupled harmonic oscillators~\cite{Alzar02, Joe06, Rodriguez16}. When the oscillators are detuned and one is much more damped than the other, interference effects lead to an asymmetric Fano-like resonance in the spectrum of the heavily damped oscillator~\cite{Joe06}. Conversely, when the two oscillators are strongly coupled, weakly damped, and tuned in resonance, the total energy is split between two new eigenmodes at different frequencies~\cite{Alzar02, Rodriguez16}. This is the so-called Rabi splitting or normal mode splitting. Within the classical framework, the main difference between the two effects sits in the ratio of the frequency detuning to the total loss rates of the oscillators. This ratio is large for a Fano resonance and close to zero for a Rabi splitting. The common aspect is the need for coherent coupling between two oscillators. Without this key ingredient, the bare oscillators simply display Lorentzian lineshapes around their resonance frequencies.

Interest in Fano resonances and Rabi splittings is so large that a complete list of references in photonics only is beyond our reach. Nonetheless, we can highlight a few scenarios where Fano resonances are relevant: sensing~\cite{Hao08, Offermans11}, switching~\cite{Mork14}, directional scattering~\cite{Nordlander10, Dionne11, Martin17}, spontaneous emission~\cite{Doeleman20}, lasing~\cite{Mork17}, and non-reciprocity~\cite{Alu20}. Meanwhile, Rabi splittings have attracted interest for enhancing or modifying chemical landscapes~\cite{Ebbesen12,Feist18}, optical nonlinearities~\cite{Rodriguez17}, electrical conductivities~\cite{Laussy10, Ebbesen15}, biological processes~\cite{Coles14}, lasing~\cite{Su17, Ramezani17}, and quantum light emission~\cite{Delteil19,Volz19}. Key to progress in all these directions is the correct identification of Fano resonances and Rabi splittings based on optical measurements. A first challenge in this endeavor arises because the transmittance ($T$), reflectance ($R$), absorbance ($A$), and photoluminescence ($PL$), of a fixed light-matter system generally display different features~\cite{Savona95}. In particular, Rabi splittings observed in $T$, $R$, $A$, and $PL$ are in general all different. The differences can be so large that some observables display a well-resolved Rabi splitting while other observables display no splitting at all. This effect has been discussed in the literature~\cite{Ebbesen11, Rakovich}, and coupled oscillator analogs can shed some light into its origin~\cite{Rodriguez12}. In this manuscript, we consider a second challenge in the identification of Fano resonances and Rabi splittings --- one that appears to have never been considered, yet is highly relevant to experiments. In particular, we ask: Can the numerical aperture(s) of the measuring instrument artificially generate or obscure Fano resonances and Rabi splittings? To address this question, we performed experiments with the simplest and most widely used optical resonator: a Fabry-P\'erot cavity. Our cavity contains a perovskite crystal of contemporary interest, namely CsPbBr\textsubscript{3}~\cite{Liu18, Xiong20}. Our coupled oscillator system is therefore one that comprises cavity photons and semiconductor excitons. As we will show, the choice of numerical aperture(s) used to probe this system conveys a number of surprises and spectral artefacts which can lead to misleading conclusions when not properly considered.

\begin{figure}[H]
	\includegraphics[width=1\linewidth]{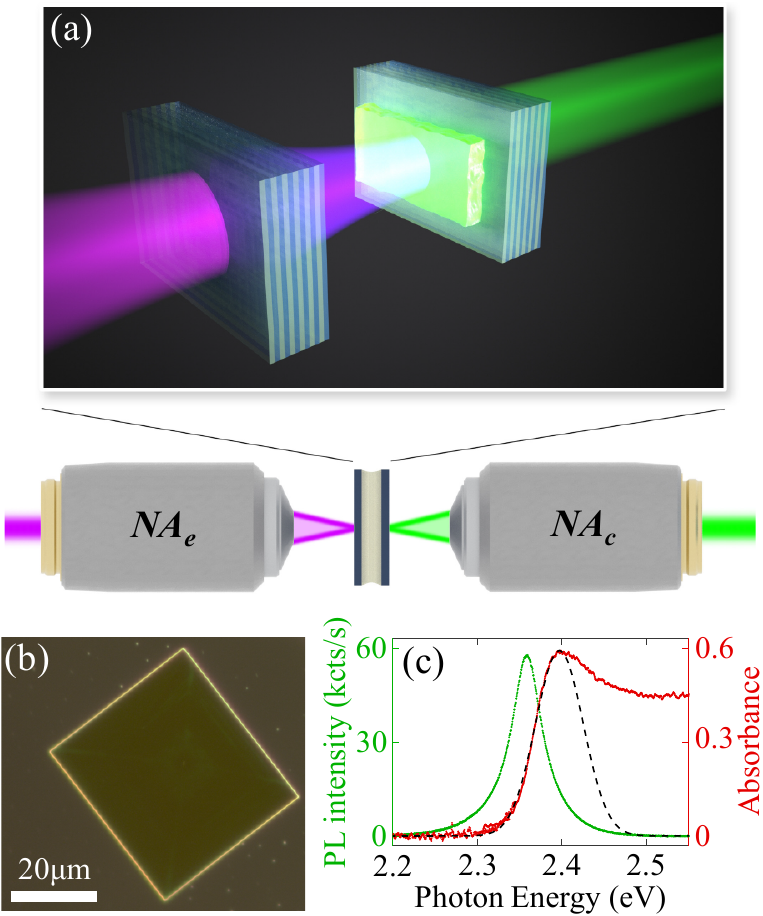}
	\caption{(a) Tunable cavity with a CsPbBr\textsubscript{3} crystal inside. The excitation light is coupled into the cavity through a microscope objective with numerical aperture $NA_e$ and light is collected through a different objective with  numerical aperture $NA_c$. (b) Dark-field microscope image of a typical CsPbBr\textsubscript{3} crystal on a glass substrate. (c) Absorbance and photoluminescence spectra of the CsPbBr\textsubscript{3} crystal used for the experiments in Figs.~\ref{fig2}, ~\ref{fig3}, and ~\ref{fig5}. The excitonic peak in the absorbance spectrum is fitted with a Gaussian distribution centered at $2.397\pm0.002$ eV and a standard deviation of 66 meV. The photoluminescence is fitted with a Lorentzian lineshape centered at $2.358\pm0.002$ eV and with a 44 meV linewidth.}
	\label{fig1}
\end{figure}

Figure~\ref{fig1}a  illustrates our experimental system: a tunable Fabry-P\'erot cavity with a CsPbBr\textsubscript{3} crystal inside. CsPbBr\textsubscript{3} crystals  were synthesized on a mica substrate via chemical vapor deposition (CVD)~\cite{Xiong16}. Using thermal release tape, the crystals were subsequently transferred onto a glass substrate for characterization, or onto a mirror for experiments.  Figure~\ref{fig1}b shows a dark-field image of a typical CsPbBr\textsubscript{3} crystal on a glass substrate. Fig.~\ref{fig1}c  shows the absorbance and $PL$  spectrum of the CsPbBr\textsubscript{3} crystal used in all our experiments discussed below. The absorbance spectrum has an excitonic peak at $2.397\pm0.002$ eV with a 66 meV linewidth, estimated via the Gaussian fit shown in Fig.~\ref{fig1}c . The $PL$ spectrum is fitted  with a Lorentzian lineshape centered at $2.358\pm0.002$ eV and with 44 meV linewidth.  We placed this crystal in our cavity, which is made by two distributed Bragg reflectors (DBRs) with a peak reflectance of $99.9$\% at $530$ nm. The position and orientation of one of the cavity mirrors are controlled with a six degree-of-freedom piezoelectric actuator. The other mirror, coated with the CsPbBr\textsubscript{3} crystal, is mounted on three piezoelectric actuators. Two of the actuators are used to place the CsPbBr\textsubscript{3} crystal along the optical axis; the other actuator serves to finely adjust the cavity length.  Using this setup, we measured the $T$ and $PL$ spectrum as a function of the cavity length. We use an incoherent white light source for $T$ measurements, and a 405 nm laser for $PL$ measurements.  Optical excitation and collection were achieved through microscope objectives with numerical aperture as specified for each figure below. We refer to the excitation $NA$ as $NA_e$, and to the collection $NA$ as $NA_c$. For $T$ measurements, we varied  $NA_e$ while keeping $NA_c=0.4$ constant. For $PL$ measurements,  we varied  $NA_c$ while keeping $NA_e=0.25$ constant. In the $PL$ case, the choice $NA_e=0.25$ is not so relevant because the incident laser is  filtered out and we only collect the luminescence. Moreover, the laser power for all $PL$ measurements is sufficiently low (40 $\mu$W at the excitation objective) to avoid nonlinear effects and crystal degradation.

\begin{figure}[!]
	\includegraphics[width=1\linewidth]{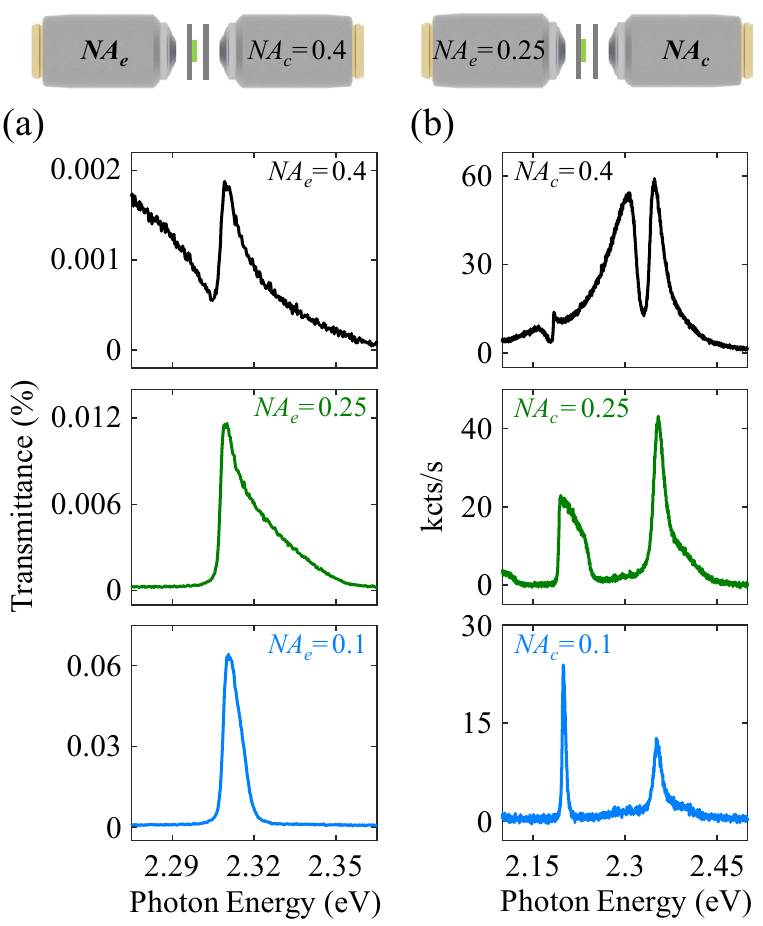}
	\caption{(a) Transmittance and (b) photoluminescence spectra of the CsPbBr\textsubscript{3}-cavity system. In (a) we vary the excitation numerical aperture  $NA_e$ while keeping the collection numerical aperture  $NA_c=0.4$ constant. In (b)  we vary $NA_c$ while keeping $NA_e=0.25$  constant.   The cavity length is $L=3440$ nm in (a), and $L=3360$ nm in (b).}
	\label{fig2}
\end{figure}

Figure~\ref{fig2}a shows $T$ spectra of the CsPbBr\textsubscript{3}-cavity system at a cavity length $L=3440$ nm, for three different $NA_e$. For $NA_e=0.1$, we observe a nearly-symmetric resonance peak on a flat background. The same optical resonance acquires a high-energy tail when  $NA_e$ increases to $0.25$. Then, an asymmetric lineshape resembling a Fano resonance appears for $NA_e=0.4$. This lineshape is not the result of Fano interference. Instead, it is an artefact of the large  $NA_e$. As demonstrated ahead, the asymmetric lineshape is due to the incoherent sum of transmitted intensities  at different angles.

Figure~\ref{fig2}b shows $PL$ spectra of the same CsPbBr\textsubscript{3}-cavity system, for a slightly different cavity length $L=3360$ nm. For $NA_c=0.1$, we observe two nearly-symmetric resonance peaks on a flat background. Each of these peaks corresponds to a cavity resonance. For $NA_c=0.25$, the peaks acquire a high-energy tail. For $NA_c=0.4$, the measured spectrum displays two remarkable features: a feature resembling a Rabi splitting appears around  2.32 eV, and a Fano-like resonance appears around 2.18 eV. These spectral features, which only appear for a sufficiently large $NA_c$, are artefacts. They are due to the incoherent sum of the cavity-enhanced perovskite emission at different angles. The apparent Rabi splitting  in Fig.~\ref{fig2}b,  around 2.33 eV, is suspiciously close to the bare exciton energy at 2.397 eV. However, this proximity is only a coincidence.  In order to elucidate the origin of all these artefacts, we proceed to inspect spectra across a wide range of cavity lengths.

\begin{figure}[!]
	\includegraphics[width=1\linewidth]{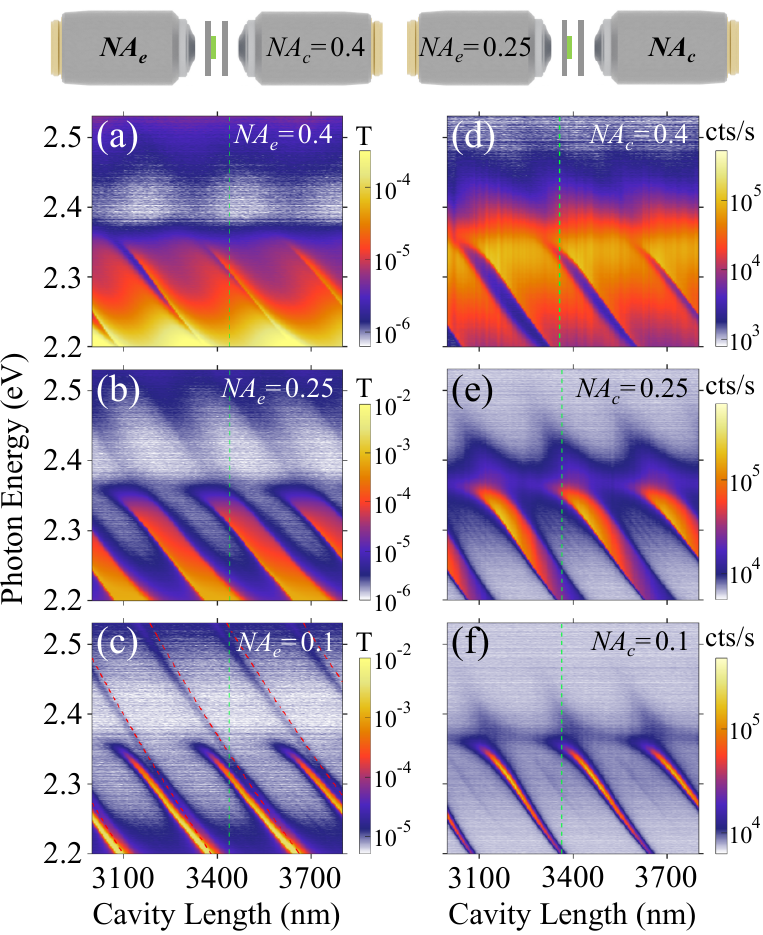}
	\caption{(a)-(c) Transmittance and (d)-(f) photoluminescence spectra of the CsPbBr\textsubscript{3}-cavity system as a function of the cavity length. In (a)-(c) we vary the excitation numerical aperture  $NA_e$ while keeping the collection numerical aperture  $NA_c=0.4$ constant. In (d)-(f) we vary $NA_c$ while keeping $NA_e=0.25$  constant.  The bright bands in all measurements correspond to resonances associated with different longitudinal mode numbers. The resonances are blurred at high energies by the bandgap absorption of CsPbBr\textsubscript{3}. For reference, (c) shows empty cavity modes as red dashed lines. Green dashed lines in (a)-(c) and (d)-(f) indicate the cavity lengths inspected in Fig.~\ref{fig2}(a) and Fig.~\ref{fig2}(b), respectively.}
	\label{fig3}
\end{figure}

Figure~\ref{fig3} shows $T$ and $PL$ spectra as a function of cavity length for the same configurations considered in  Fig.~\ref{fig2}. For reference, the vertical green dashed lines in Figs.~\ref{fig3}a-c (resp. Figs.~\ref{fig3}d-f) indicate the cavity length considered in Fig.~\ref{fig2}a (resp. Fig.~\ref{fig2}b).    Each bright band in the color plot corresponds to a resonance associated with a particular longitudinal mode of the cavity. The longitudinal modes of the empty cavity are shown as  red dashed lines  in Fig.~\ref{fig3}c. Their resonance frequency satisfies  $f = qc  / 2L$, with $q$ the longitudinal mode number, $c$ the speed of light, and $L$ the cavity length.

We first analyze the $T$ spectra in  Figs.~\ref{fig3}a-c. Notice how, as  $NA_e$ increases, all resonances broaden in energy. For $NA_e=0.4$, shown in Fig.~\ref{fig3}a, the transmission from consecutive longitudinal modes nearly overlap. Only a narrow transmission dip between resonances remains. This is the Fano-like dip observed in  Fig.~\ref{fig2}a.  The results in Figs.~\ref{fig3}a-c  already reveal the origin of the transmission dip: as $NA_e$ increases, transmission bands associated with consecutive $q$ increasingly approach each other in energy and eventually overlap. As Figs.~\ref{fig3}d-f  show, similar behavior arises in the $PL$ measurements when  $NA_c$ increases.  For $NA_c=0.4$, shown in Fig.~\ref{fig3}d, the emission from   consecutive longitudinal modes nearly overlap in energy.  The  emission dip between bands with consecutive $q$, at fixed cavity length, results in features that can resemble Fano resonances or Rabi splittings but are neither.

Our measurements clearly demonstrate that even a moderately large $NA$ can generate spectral features reminiscent of coherently coupled systems. Two important questions remain. First, are these spectral features due to the presence of an excitonic material in our cavity? Second, are interference effects responsible for these spectral features in any way? In the following, we will demonstrate that the answer to both of these questions is negative. We will analyze $T$ spectra for an empty Fabry-P\'erot cavity, in experiments and theory. While this analysis  is necessarily restricted to $T$ (there is no $PL$ in an empty cavity), the insights obtained from it are general.

\begin{figure}[!]
	\includegraphics[width=1\linewidth]{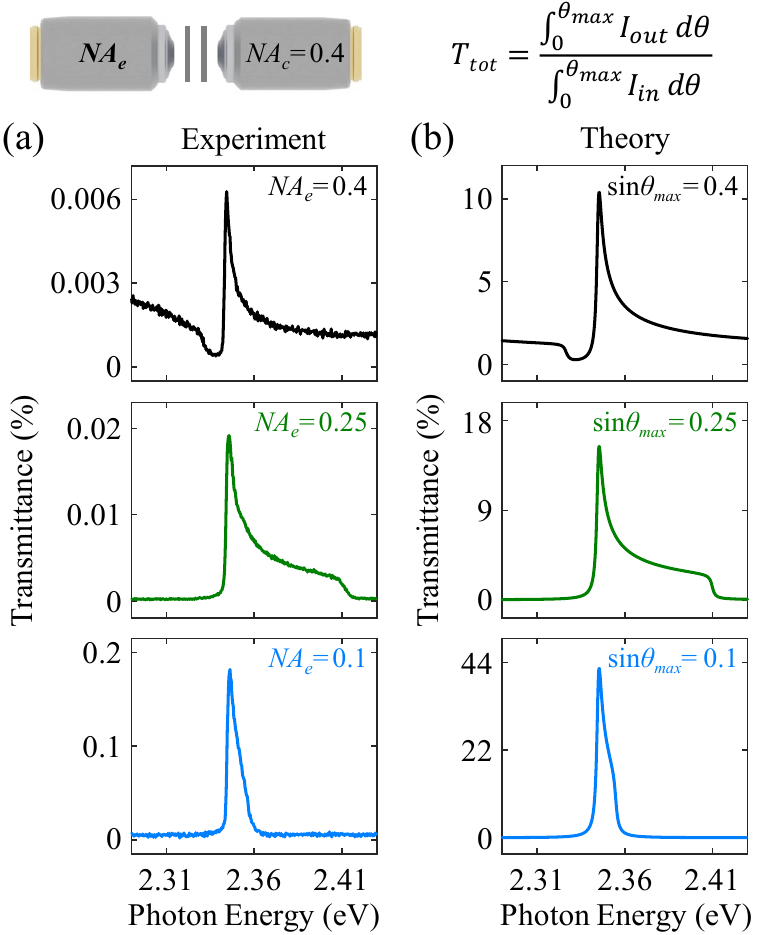}
	\caption{(a) Experimental transmittance spectra of an empty cavity for three different $NA_e$. (b) Transfer matrix calculations of  transmittance spectra averaged over an angular range corresponding to the experimental $NA_e$. The averaged transmittance is given by the incoherent sum of transmitted intensities at different angles, divided by the incident intensity over the same angular range. The cavity length is $L = 3430$ nm in both experiments and calculations.}
	\label{fig4}
\end{figure}

Figure~\ref{fig4}a shows experimental $T$ spectra for the same three $NA_e$ considered in Figs.~\ref{fig2} and~\ref{fig3}, and at a cavity length $L=3430$ nm. As $NA_e$ increases, we again observe a broadening of the resonance and the emergence of a Fano-like lineshape. This demonstrates how increasing $NA_e$ generates  a Fano-like lineshape even in the absence of an excitonic material. To identify the mechanism underlying this effect, we use a transfer matrix model to calculate the transmission of an empty cavity. Our goal is to determine whether the incoherent sum of transmitted intensities over a finite angular range can lead to Fano-like lineshapes.

Our transfer matrix calculations are done for a DBR-vacuum-DBR cavity. We model each DBR as a stack of 6 pairs of layers with refractive index $n_1 = 1.45$ and $n_1 = 2.3$. These values correspond, within the frequency range of interest, to our experimental DBRs made of silica and Ta$_2$O$_5$.  We calculate the transmittance (i.e., transmitted power normalized to the incident power) when  the cavity is illuminated by a single plane wave, and we vary the angle of incidence in steps of 0.005 degrees. Then, we average the transmittance over the angular range corresponding to the  $NA_e$ of interest. We also sum the contributions of the two orthogonal polarizations, in correspondence to our experiments which were done using unpolarized light.

Figure~\ref{fig4}b shows transfer matrix calculations results obtained as described above, for the same cavity length considered in Fig.~\ref{fig4}a. Notice how the experimentally-observed Fano-like lineshapes are reproduced in the calculations. This demonstrates that the sum of transmitted intensities at different incident angles suffices to generate Fano-like lineshapes. Since we are adding intensities and not field amplitudes, we are neglecting interference effects due to different plane waves. Hence, by reproducing the experimentally-observed lineshapes in this way, we can conclude that the measured lineshapes are not due to Fano interference. We therefore call these Fano-like lineshapes `artefacts'. While our calculations qualitatively reproduce the experimental lineshapes very well, the value of $T$ is much lower in experiments. We attribute this difference to an imperfect alignment of the cavity mirrors,  which leads to very significant optical losses through the sides of the cavity. Scattering losses may also reduce the value of $T$ in experiments, but we suspect this contribution is smaller given the uncertainty we have in the angular alignment ($\sim 5$ millidegrees).

\begin{figure}[!]
	\includegraphics[width=1\linewidth]{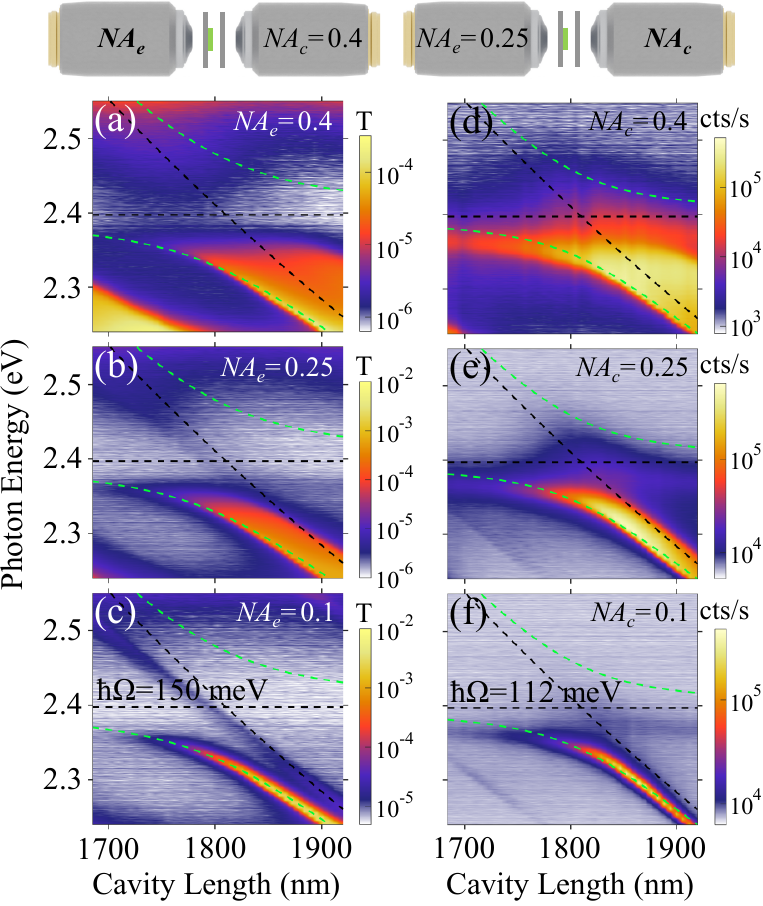}
	\caption{(a)-(c) Transmittance  and (d)-(f) photoluminescence spectra of the CsPbBr\textsubscript{3}-cavity system as a function of the cavity length, for a shorter cavity than in Figs.~\ref{fig2} and~\ref{fig3}. The horizontal dashed black lines indicate the exciton energy, and the tilted dashed black lines indicate the energy of the $q=7$ longitudinal cavity mode. The green dashed lines in all panels are the eigenvalues of a $2\times 2$ Hamiltonian representing the exciton-photon coupled system. The coupling constant is the only fit parameter, and the result is shown in (c) and (f) for transmittance and photoluminescence measurements, respectively. The coupling was not changed for measurements with different $NA$.  }
	\label{fig5}
\end{figure}

Having shown that the $NA$ can generate spectral lineshapes resembling those characterizing coherently coupled systems, we turn our attention to the complementary question: Can the $NA$ obscure spectral features of bona fide coherently coupled systems?  To answer this question, we inspect  $T$ and $PL$ spectra of our perovskite-cavity system for shorter cavity lengths than considered in Figs.~\ref{fig2} and~\ref{fig3}. Figure~\ref{fig5}a-c show measurements in a parameter range where the $q=7$ cavity mode crosses the exciton energy. The energies of the bare cavity mode and exciton are shown as dashed black lines in all panels in Fig.~\ref{fig5}.

Let us first consider the results for the smallest $NA_e$ shown in Fig.~\ref{fig5}c. Notice how, as the cavity length decreases,  the resonance peak bends away from the empty cavity mode and does not cross the exciton energy (2.397 eV).  This anti-crossing behavior is characteristic of strong coupling. To estimate the coupling regime (weak vs strong), we fit the measured dispersion with the eigenvalues of a $2 \times 2$ Hamiltonian describing our exciton-photon coupled system. The diagonal terms of the Hamiltonian contain the exciton and cavity photon energies. We know the bare exciton energy from $A$ measurements in Fig.~\ref{fig1}c, and the bare cavity photon energy from the relation $f = qc  / 2L$.  The off-diagonal term of the Hamiltonian, i.e. the coupling strength, is the only fit parameter in our model. In this way, we estimate a Rabi splitting of 150 meV as shown in Fig.~\ref{fig5}c. This value is well above the sum of the exciton and cavity photon linewidths, which are 66 meV and 0.5 meV, respectively. Therefore, our exciton-photon system is in the strong coupling regime and the observed resonances correspond to exciton-polaritons. This is not a surprising result. Indeed, strong exciton-photon coupling was recently shown in similar Fabry-P\'erot cavities filled with the same perovskite semiconductor.~\cite{Liu18, Xiong20}. The interesting new observation, enabled by our tunable cavity system, is that the polariton band is so broadened for $NA_e=0.4$  that its central energy appears to coincide with the bare cavity mode. Measuring spectra with a single moderately large $NA_e$ only, as done in many works, can lead to the erroneous conclusion that there is no strong coupling. The same holds for the $PL$ measurements shown in Figs.~\ref{fig5}d-f. These results clearly demonstrate how a moderately large  $NA$ can obscure the spectral signature of a bona fide strongly coupled system.

The measurements in Fig.~\ref{fig5} display features which, while not crucial to our analysis, require clarification. First, the upper polariton is not visible in Fig.~\ref{fig5} because of re-absorption. This can be understood in light of the absorbance measurements in Fig.~\ref{fig1}c. There one can see that absorbance is much greater at energies above the exciton energy (where the upper polariton resides) than below. Indeed, previous works reporting exciton-polaritons in optical cavities filled with the same perovskite semiconductor also did not observe the upper polariton~\cite{Xiong18}.  Second, the Rabi splitting observed in $PL$ (112 meV) is smaller than the one observed in $T$ (150 meV).  This is consistent with previous calculations~\cite{Savona95} and many experimental observations of polariton systems~\cite{Ebbesen11}.

In conclusion, we have shown that a moderately large numerical aperture can artificially generate and obscure spectral features of coupled light-matter systems. In particular, we have seen Fano-like resonances and apparent Rabi splittings entirely due to the incoherent sum of transmitted intensities at different angles collected by the $NA$. Moreover, we have seen how the spectral signature  of a bona fide strongly coupled system --- the Rabi splitting --- can be obscured in measurements with a large $NA$.  While these results were obtained using a tunable Fabry-P\'erot cavity, we believe that the spectral artefacts we report can also be found in other dispersive nanophotonic systems of contemporary interest, such as plasmonic gratings and metasurfaces.  In general, Fano-like lineshapes can be artificially generated whenever the resonance frequency of an optical mode is not constant across the angular range of the measurement. Meanwhile, spectral features resembling Rabi splittings can emerge whenever the frequency separation between two (orthogonal) modes is smaller than the apparent linewidths which are artificially broadenend by the numerical aperture.  These spectral  artefacts can be avoided by measuring with the smallest possible $NA$, such that the resonance frequency is constant over the limited angular range of the $NA$.  Alternatively, angle-resolved measurements can also avoid pitfalls.  We recognize that these recommendations may be difficult to implement when measuring the transmission of small samples, as often the case in nanophotonics research. In that case, a full energy-angle-resolved theoretical study reproducing the measured lineshapes with good accuracy seems to be the only way to assess the true nature of Fano-like lineshapes and Rabi-like splittings. In any case, the mere experimental observation of spectral lineshapes resembling Fano resonances or Rabi splittings cannot by itself be taken as solid evidence of coherent optical phenomena.

\section*{Acknowledgments}
\noindent This work is part of the research programme of the Netherlands Organisation for Scientific Research (NWO). We thank Femius Koenderink for stimulating discussions.  S.R.K.R. acknowledges an ERC Starting Grant with project number 852694. \\


\begin{thebibliography}{33}%
\makeatletter
\providecommand \@ifxundefined [1]{%
 \@ifx{#1\undefined}
}%
\providecommand \@ifnum [1]{%
 \ifnum #1\expandafter \@firstoftwo
 \else \expandafter \@secondoftwo
 \fi
}%
\providecommand \@ifx [1]{%
 \ifx #1\expandafter \@firstoftwo
 \else \expandafter \@secondoftwo
 \fi
}%
\providecommand \natexlab [1]{#1}%
\providecommand \enquote  [1]{``#1''}%
\providecommand \bibnamefont  [1]{#1}%
\providecommand \bibfnamefont [1]{#1}%
\providecommand \citenamefont [1]{#1}%
\providecommand \href@noop [0]{\@secondoftwo}%
\providecommand \href [0]{\begingroup \@sanitize@url \@href}%
\providecommand \@href[1]{\@@startlink{#1}\@@href}%
\providecommand \@@href[1]{\endgroup#1\@@endlink}%
\providecommand \@sanitize@url [0]{\catcode `\\12\catcode `\$12\catcode
  `\&12\catcode `\#12\catcode `\^12\catcode `\_12\catcode `\%12\relax}%
\providecommand \@@startlink[1]{}%
\providecommand \@@endlink[0]{}%
\providecommand \url  [0]{\begingroup\@sanitize@url \@url }%
\providecommand \@url [1]{\endgroup\@href {#1}{\urlprefix }}%
\providecommand \urlprefix  [0]{URL }%
\providecommand \Eprint [0]{\href }%
\providecommand \doibase [0]{https://doi.org/}%
\providecommand \selectlanguage [0]{\@gobble}%
\providecommand \bibinfo  [0]{\@secondoftwo}%
\providecommand \bibfield  [0]{\@secondoftwo}%
\providecommand \translation [1]{[#1]}%
\providecommand \BibitemOpen [0]{}%
\providecommand \bibitemStop [0]{}%
\providecommand \bibitemNoStop [0]{.\EOS\space}%
\providecommand \EOS [0]{\spacefactor3000\relax}%
\providecommand \BibitemShut  [1]{\csname bibitem#1\endcsname}%
\let\auto@bib@innerbib\@empty
\bibitem [{\citenamefont {Miroshnichenko}\ \emph {et~al.}(2010)\citenamefont
  {Miroshnichenko}, \citenamefont {Flach},\ and\ \citenamefont
  {Kivshar}}]{Miroshnichenko10}%
  \BibitemOpen
  \bibfield  {author} {\bibinfo {author} {\bibfnamefont {A.~E.}\ \bibnamefont
  {Miroshnichenko}}, \bibinfo {author} {\bibfnamefont {S.}~\bibnamefont
  {Flach}},\ and\ \bibinfo {author} {\bibfnamefont {Y.~S.}\ \bibnamefont
  {Kivshar}},\ }\bibfield  {title} {\bibinfo {title} {{F}ano resonances in
  nanoscale structures},\ }\href {https://doi.org/10.1103/RevModPhys.82.2257}
  {\bibfield  {journal} {\bibinfo  {journal} {Rev. Mod. Phys.}\ }\textbf
  {\bibinfo {volume} {82}},\ \bibinfo {pages} {2257} (\bibinfo {year}
  {2010})}\BibitemShut {NoStop}%
\bibitem [{\citenamefont {Luk'yanchuk}\ \emph {et~al.}(2010)\citenamefont
  {Luk'yanchuk}, \citenamefont {Zheludev}, \citenamefont {Maier}, \citenamefont
  {Halas}, \citenamefont {Nordlander}, \citenamefont {Giessen},\ and\
  \citenamefont {Chong}}]{Nordlander10}%
  \BibitemOpen
  \bibfield  {author} {\bibinfo {author} {\bibfnamefont {B.}~\bibnamefont
  {Luk'yanchuk}}, \bibinfo {author} {\bibfnamefont {N.~I.}\ \bibnamefont
  {Zheludev}}, \bibinfo {author} {\bibfnamefont {S.~A.}\ \bibnamefont {Maier}},
  \bibinfo {author} {\bibfnamefont {N.~J.}\ \bibnamefont {Halas}}, \bibinfo
  {author} {\bibfnamefont {P.}~\bibnamefont {Nordlander}}, \bibinfo {author}
  {\bibfnamefont {H.}~\bibnamefont {Giessen}},\ and\ \bibinfo {author}
  {\bibfnamefont {C.~T.}\ \bibnamefont {Chong}},\ }\bibfield  {title} {\bibinfo
  {title} {{T}he {F}ano resonance in plasmonic nanostructures and
  metamaterials},\ }\href {https://www.nature.com/articles/nmat2810} {\bibfield
   {journal} {\bibinfo  {journal} {Nat. Mater.}\ }\textbf {\bibinfo {volume}
  {9}},\ \bibinfo {pages} {707} (\bibinfo {year} {2010})}\BibitemShut {NoStop}%
\bibitem [{\citenamefont {T\"{o}rma}\ and\ \citenamefont
  {Barnes}(2014)}]{Torma14}%
  \BibitemOpen
  \bibfield  {author} {\bibinfo {author} {\bibfnamefont {P.}~\bibnamefont
  {T\"{o}rma}}\ and\ \bibinfo {author} {\bibfnamefont {W.~L.}\ \bibnamefont
  {Barnes}},\ }\bibfield  {title} {\bibinfo {title} {Strong coupling between
  surface plasmon polaritons and emitters: {A} review},\ }\href
  {https://doi.org/10.1088/0034-4885/78/1/013901} {\bibfield  {journal}
  {\bibinfo  {journal} {Rep. Prog. Phys.}\ }\textbf {\bibinfo {volume} {78}},\
  \bibinfo {pages} {013901} (\bibinfo {year} {2014})}\BibitemShut {NoStop}%
\bibitem [{\citenamefont {Limonov}\ \emph {et~al.}(2017)\citenamefont
  {Limonov}, \citenamefont {Rybin}, \citenamefont {Poddubny},\ and\
  \citenamefont {Kivshar}}]{Rybin17}%
  \BibitemOpen
  \bibfield  {author} {\bibinfo {author} {\bibfnamefont {M.~F.}\ \bibnamefont
  {Limonov}}, \bibinfo {author} {\bibfnamefont {M.~V.}\ \bibnamefont {Rybin}},
  \bibinfo {author} {\bibfnamefont {A.~N.}\ \bibnamefont {Poddubny}},\ and\
  \bibinfo {author} {\bibfnamefont {Y.~S.}\ \bibnamefont {Kivshar}},\
  }\bibfield  {title} {\bibinfo {title} {{F}ano resonances in photonics},\
  }\href {https://www.nature.com/articles/nphoton.2017.142} {\bibfield
  {journal} {\bibinfo  {journal} {Nat. Photonics}\ }\textbf {\bibinfo {volume}
  {11}},\ \bibinfo {pages} {543} (\bibinfo {year} {2017})}\BibitemShut
  {NoStop}%
\bibitem [{\citenamefont {Garrido~Alzar}\ \emph {et~al.}(2002)\citenamefont
  {Garrido~Alzar}, \citenamefont {Martinez},\ and\ \citenamefont
  {Nussenzveig}}]{Alzar02}%
  \BibitemOpen
  \bibfield  {author} {\bibinfo {author} {\bibfnamefont {C.}~\bibnamefont
  {Garrido~Alzar}}, \bibinfo {author} {\bibfnamefont {M.}~\bibnamefont
  {Martinez}},\ and\ \bibinfo {author} {\bibfnamefont {P.}~\bibnamefont
  {Nussenzveig}},\ }\bibfield  {title} {\bibinfo {title} {Classical analog of
  electromagnetically induced transparency},\ }\href
  {https://doi.org/10.1119/1.1412644} {\bibfield  {journal} {\bibinfo
  {journal} {Am. J. Phys.}\ }\textbf {\bibinfo {volume} {70}},\ \bibinfo
  {pages} {37} (\bibinfo {year} {2002})}\BibitemShut {NoStop}%
\bibitem [{\citenamefont {Joe}\ \emph {et~al.}(2006)\citenamefont {Joe},
  \citenamefont {Satanin},\ and\ \citenamefont {Kim}}]{Joe06}%
  \BibitemOpen
  \bibfield  {author} {\bibinfo {author} {\bibfnamefont {Y.~S.}\ \bibnamefont
  {Joe}}, \bibinfo {author} {\bibfnamefont {A.~M.}\ \bibnamefont {Satanin}},\
  and\ \bibinfo {author} {\bibfnamefont {C.~S.}\ \bibnamefont {Kim}},\
  }\bibfield  {title} {\bibinfo {title} {Classical analogy of {F}ano
  resonances},\ }\href {https://doi.org/10.1088/0031-8949/74/2/020} {\bibfield
  {journal} {\bibinfo  {journal} {Phys. Scr.}\ }\textbf {\bibinfo {volume}
  {74}},\ \bibinfo {pages} {259} (\bibinfo {year} {2006})}\BibitemShut
  {NoStop}%
\bibitem [{\citenamefont {Rodriguez}(2016)}]{Rodriguez16}%
  \BibitemOpen
  \bibfield  {author} {\bibinfo {author} {\bibfnamefont {S.~R.-K.}\
  \bibnamefont {Rodriguez}},\ }\bibfield  {title} {\bibinfo {title}
  {{C}lassical and quantum distinctions between weak and strong coupling},\
  }\href {https://iopscience.iop.org/article/10.1088/0143-0807/37/2/025802}
  {\bibfield  {journal} {\bibinfo  {journal} {Eur. J. Phys.}\ }\textbf
  {\bibinfo {volume} {37}},\ \bibinfo {pages} {025802} (\bibinfo {year}
  {2016})}\BibitemShut {NoStop}%
\bibitem [{\citenamefont {Hao}\ \emph {et~al.}(2008)\citenamefont {Hao},
  \citenamefont {Sonnefraud}, \citenamefont {Dorpe}, \citenamefont {Maier},
  \citenamefont {Halas},\ and\ \citenamefont {Nordlander}}]{Hao08}%
  \BibitemOpen
  \bibfield  {author} {\bibinfo {author} {\bibfnamefont {F.}~\bibnamefont
  {Hao}}, \bibinfo {author} {\bibfnamefont {Y.}~\bibnamefont {Sonnefraud}},
  \bibinfo {author} {\bibfnamefont {P.~V.}\ \bibnamefont {Dorpe}}, \bibinfo
  {author} {\bibfnamefont {S.~A.}\ \bibnamefont {Maier}}, \bibinfo {author}
  {\bibfnamefont {N.~J.}\ \bibnamefont {Halas}},\ and\ \bibinfo {author}
  {\bibfnamefont {P.}~\bibnamefont {Nordlander}},\ }\bibfield  {title}
  {\bibinfo {title} {Symmetry breaking in plasmonic nanocavities: Subradiant
  {LSPR} sensing and a tunable {F}ano resonance},\ }\href
  {https://doi.org/10.1021/nl802509r} {\bibfield  {journal} {\bibinfo
  {journal} {Nano Lett.}\ }\textbf {\bibinfo {volume} {8}},\ \bibinfo {pages}
  {3983} (\bibinfo {year} {2008})}\BibitemShut {NoStop}%
\bibitem [{\citenamefont {Offermans}\ \emph {et~al.}(2011)\citenamefont
  {Offermans}, \citenamefont {Schaafsma}, \citenamefont {Rodriguez},
  \citenamefont {Zhang}, \citenamefont {Crego-Calama}, \citenamefont
  {Brongersma},\ and\ \citenamefont {Gómez~Rivas}}]{Offermans11}%
  \BibitemOpen
  \bibfield  {author} {\bibinfo {author} {\bibfnamefont {P.}~\bibnamefont
  {Offermans}}, \bibinfo {author} {\bibfnamefont {M.~C.}\ \bibnamefont
  {Schaafsma}}, \bibinfo {author} {\bibfnamefont {S.~R.~K.}\ \bibnamefont
  {Rodriguez}}, \bibinfo {author} {\bibfnamefont {Y.}~\bibnamefont {Zhang}},
  \bibinfo {author} {\bibfnamefont {M.}~\bibnamefont {Crego-Calama}}, \bibinfo
  {author} {\bibfnamefont {S.~H.}\ \bibnamefont {Brongersma}},\ and\ \bibinfo
  {author} {\bibfnamefont {J.}~\bibnamefont {Gómez~Rivas}},\ }\bibfield
  {title} {\bibinfo {title} {Universal {S}caling of the {F}igure of {M}erit of
  {P}lasmonic {S}ensors},\ }\href {https://doi.org/10.1021/nn201227b}
  {\bibfield  {journal} {\bibinfo  {journal} {ACS Nano}\ }\textbf {\bibinfo
  {volume} {5}},\ \bibinfo {pages} {5151} (\bibinfo {year} {2011})}\BibitemShut
  {NoStop}%
\bibitem [{\citenamefont {Yu}\ \emph {et~al.}(2014)\citenamefont {Yu},
  \citenamefont {Heuck}, \citenamefont {Hu}, \citenamefont {Xue}, \citenamefont
  {Peucheret}, \citenamefont {Chen}, \citenamefont {Oxenløwe}, \citenamefont
  {Yvind},\ and\ \citenamefont {Mørk}}]{Mork14}%
  \BibitemOpen
  \bibfield  {author} {\bibinfo {author} {\bibfnamefont {Y.}~\bibnamefont
  {Yu}}, \bibinfo {author} {\bibfnamefont {M.}~\bibnamefont {Heuck}}, \bibinfo
  {author} {\bibfnamefont {H.}~\bibnamefont {Hu}}, \bibinfo {author}
  {\bibfnamefont {W.}~\bibnamefont {Xue}}, \bibinfo {author} {\bibfnamefont
  {C.}~\bibnamefont {Peucheret}}, \bibinfo {author} {\bibfnamefont
  {Y.}~\bibnamefont {Chen}}, \bibinfo {author} {\bibfnamefont {L.~K.}\
  \bibnamefont {Oxenløwe}}, \bibinfo {author} {\bibfnamefont {K.}~\bibnamefont
  {Yvind}},\ and\ \bibinfo {author} {\bibfnamefont {J.}~\bibnamefont {Mørk}},\
  }\bibfield  {title} {\bibinfo {title} {{F}ano resonance control in a photonic
  crystal structure and its application to ultrafast switching},\ }\href
  {https://doi.org/10.1063/1.4893451} {\bibfield  {journal} {\bibinfo
  {journal} {Appl. Phys. Lett.}\ }\textbf {\bibinfo {volume} {105}},\ \bibinfo
  {pages} {061117} (\bibinfo {year} {2014})}\BibitemShut {NoStop}%
\bibitem [{\citenamefont {Sheikholeslami}\ \emph {et~al.}(2011)\citenamefont
  {Sheikholeslami}, \citenamefont {García-Etxarri},\ and\ \citenamefont
  {Dionne}}]{Dionne11}%
  \BibitemOpen
  \bibfield  {author} {\bibinfo {author} {\bibfnamefont {S.~N.}\ \bibnamefont
  {Sheikholeslami}}, \bibinfo {author} {\bibfnamefont {A.}~\bibnamefont
  {García-Etxarri}},\ and\ \bibinfo {author} {\bibfnamefont {J.~A.}\
  \bibnamefont {Dionne}},\ }\bibfield  {title} {\bibinfo {title} {{C}ontrolling
  the {I}nterplay of {E}lectric and {M}agnetic {M}odes via {F}ano-like
  {P}lasmon {R}esonances},\ }\href {https://doi.org/10.1021/nl202143j}
  {\bibfield  {journal} {\bibinfo  {journal} {Nano Lett.}\ }\textbf {\bibinfo
  {volume} {11}},\ \bibinfo {pages} {3927} (\bibinfo {year}
  {2011})}\BibitemShut {NoStop}%
\bibitem [{\citenamefont {Yan}\ \emph {et~al.}(2017)\citenamefont {Yan},
  \citenamefont {Yang},\ and\ \citenamefont {Martin}}]{Martin17}%
  \BibitemOpen
  \bibfield  {author} {\bibinfo {author} {\bibfnamefont {C.}~\bibnamefont
  {Yan}}, \bibinfo {author} {\bibfnamefont {K.-Y.}\ \bibnamefont {Yang}},\ and\
  \bibinfo {author} {\bibfnamefont {O.~J.}\ \bibnamefont {Martin}},\ }\bibfield
   {title} {\bibinfo {title} {{F}ano-resonance-assisted metasurface for color
  routing},\ }\href {https://www.nature.com/articles/lsa201717} {\bibfield
  {journal} {\bibinfo  {journal} {Light Sci. Appl.}\ }\textbf {\bibinfo
  {volume} {6}},\ \bibinfo {pages} {e17017} (\bibinfo {year}
  {2017})}\BibitemShut {NoStop}%
\bibitem [{\citenamefont {Doeleman}\ \emph {et~al.}(2020)\citenamefont
  {Doeleman}, \citenamefont {Dieleman}, \citenamefont {Mennes}, \citenamefont
  {Ehrler},\ and\ \citenamefont {Koenderink}}]{Doeleman20}%
  \BibitemOpen
  \bibfield  {author} {\bibinfo {author} {\bibfnamefont {H.~M.}\ \bibnamefont
  {Doeleman}}, \bibinfo {author} {\bibfnamefont {C.~D.}\ \bibnamefont
  {Dieleman}}, \bibinfo {author} {\bibfnamefont {C.}~\bibnamefont {Mennes}},
  \bibinfo {author} {\bibfnamefont {B.}~\bibnamefont {Ehrler}},\ and\ \bibinfo
  {author} {\bibfnamefont {A.~F.}\ \bibnamefont {Koenderink}},\ }\bibfield
  {title} {\bibinfo {title} {{O}bservation of {C}ooperative {P}urcell
  {E}nhancements in {A}ntenna–{C}avity {H}ybrids},\ }\href
  {https://doi.org/10.1021/acsnano.0c05233} {\bibfield  {journal} {\bibinfo
  {journal} {ACS Nano}\ }\textbf {\bibinfo {volume} {14}},\ \bibinfo {pages}
  {12027} (\bibinfo {year} {2020})}\BibitemShut {NoStop}%
\bibitem [{\citenamefont {Yu}\ \emph {et~al.}(2017)\citenamefont {Yu},
  \citenamefont {Xue}, \citenamefont {Semenova}, \citenamefont {Yvind},\ and\
  \citenamefont {Mork}}]{Mork17}%
  \BibitemOpen
  \bibfield  {author} {\bibinfo {author} {\bibfnamefont {Y.}~\bibnamefont
  {Yu}}, \bibinfo {author} {\bibfnamefont {W.}~\bibnamefont {Xue}}, \bibinfo
  {author} {\bibfnamefont {E.}~\bibnamefont {Semenova}}, \bibinfo {author}
  {\bibfnamefont {K.}~\bibnamefont {Yvind}},\ and\ \bibinfo {author}
  {\bibfnamefont {J.}~\bibnamefont {Mork}},\ }\bibfield  {title} {\bibinfo
  {title} {Demonstration of a self-pulsing photonic crystal fano laser},\
  }\href {https://www.nature.com/articles/nphoton.2016.248} {\bibfield
  {journal} {\bibinfo  {journal} {Nat. Photonics}\ }\textbf {\bibinfo {volume}
  {11}},\ \bibinfo {pages} {81} (\bibinfo {year} {2017})}\BibitemShut {NoStop}%
\bibitem [{\citenamefont {Yang}\ \emph {et~al.}(2020)\citenamefont {Yang},
  \citenamefont {Skarda}, \citenamefont {Cotrufo}, \citenamefont {Dutt},
  \citenamefont {Ahn}, \citenamefont {Sawaby}, \citenamefont {Vercruysse},
  \citenamefont {Arbabian}, \citenamefont {Fan}, \citenamefont {Al{\`u}},\ and\
  \citenamefont {Vu\ifmmode \check{c}\else
  \v{c}\fi{}kovi\ifmmode~\acute{c}\else \'{c}\fi{}}}]{Alu20}%
  \BibitemOpen
  \bibfield  {author} {\bibinfo {author} {\bibfnamefont {K.~Y.}\ \bibnamefont
  {Yang}}, \bibinfo {author} {\bibfnamefont {J.}~\bibnamefont {Skarda}},
  \bibinfo {author} {\bibfnamefont {M.}~\bibnamefont {Cotrufo}}, \bibinfo
  {author} {\bibfnamefont {A.}~\bibnamefont {Dutt}}, \bibinfo {author}
  {\bibfnamefont {G.~H.}\ \bibnamefont {Ahn}}, \bibinfo {author} {\bibfnamefont
  {M.}~\bibnamefont {Sawaby}}, \bibinfo {author} {\bibfnamefont
  {D.}~\bibnamefont {Vercruysse}}, \bibinfo {author} {\bibfnamefont
  {A.}~\bibnamefont {Arbabian}}, \bibinfo {author} {\bibfnamefont
  {S.}~\bibnamefont {Fan}}, \bibinfo {author} {\bibfnamefont {A.}~\bibnamefont
  {Al{\`u}}},\ and\ \bibinfo {author} {\bibfnamefont {J.}~\bibnamefont
  {Vu\ifmmode \check{c}\else \v{c}\fi{}kovi\ifmmode~\acute{c}\else
  \'{c}\fi{}}},\ }\bibfield  {title} {\bibinfo {title} {Inverse-designed
  non-reciprocal pulse router for chip-based {L}i{DAR}},\ }\href
  {https://www.nature.com/articles/s41566-020-0606-0} {\bibfield  {journal}
  {\bibinfo  {journal} {Nat. Photonics}\ }\textbf {\bibinfo {volume} {14}},\
  \bibinfo {pages} {369} (\bibinfo {year} {2020})}\BibitemShut {NoStop}%
\bibitem [{\citenamefont {Hutchison}\ \emph {et~al.}(2012)\citenamefont
  {Hutchison}, \citenamefont {Schwartz}, \citenamefont {Genet}, \citenamefont
  {Devaux},\ and\ \citenamefont {Ebbesen}}]{Ebbesen12}%
  \BibitemOpen
  \bibfield  {author} {\bibinfo {author} {\bibfnamefont {J.~A.}\ \bibnamefont
  {Hutchison}}, \bibinfo {author} {\bibfnamefont {T.}~\bibnamefont {Schwartz}},
  \bibinfo {author} {\bibfnamefont {C.}~\bibnamefont {Genet}}, \bibinfo
  {author} {\bibfnamefont {E.}~\bibnamefont {Devaux}},\ and\ \bibinfo {author}
  {\bibfnamefont {T.~W.}\ \bibnamefont {Ebbesen}},\ }\bibfield  {title}
  {\bibinfo {title} {{M}odifying {C}hemical {L}andscapes by {C}oupling to
  {V}acuum {F}ields},\ }\href {https://doi.org/10.1002/anie.201107033}
  {\bibfield  {journal} {\bibinfo  {journal} {Angew. Chem. Int. Ed}\ }\textbf
  {\bibinfo {volume} {51}},\ \bibinfo {pages} {1592} (\bibinfo {year}
  {2012})}\BibitemShut {NoStop}%
\bibitem [{\citenamefont {Feist}\ \emph {et~al.}(2018)\citenamefont {Feist},
  \citenamefont {Galego},\ and\ \citenamefont {Garcia-Vidal}}]{Feist18}%
  \BibitemOpen
  \bibfield  {author} {\bibinfo {author} {\bibfnamefont {J.}~\bibnamefont
  {Feist}}, \bibinfo {author} {\bibfnamefont {J.}~\bibnamefont {Galego}},\ and\
  \bibinfo {author} {\bibfnamefont {F.~J.}\ \bibnamefont {Garcia-Vidal}},\
  }\bibfield  {title} {\bibinfo {title} {Polaritonic chemistry with organic
  molecules},\ }\href {https://doi.org/10.1021/acsphotonics.7b00680} {\bibfield
   {journal} {\bibinfo  {journal} {ACS Photonics}\ }\textbf {\bibinfo {volume}
  {5}},\ \bibinfo {pages} {205} (\bibinfo {year} {2018})}\BibitemShut {NoStop}%
\bibitem [{\citenamefont {Rodriguez}\ \emph {et~al.}(2017)\citenamefont
  {Rodriguez}, \citenamefont {Casteels}, \citenamefont {Storme}, \citenamefont
  {Carlon~Zambon}, \citenamefont {Sagnes}, \citenamefont {Le~Gratiet},
  \citenamefont {Galopin}, \citenamefont {Lema\^{\i}tre}, \citenamefont {Amo},
  \citenamefont {Ciuti},\ and\ \citenamefont {Bloch}}]{Rodriguez17}%
  \BibitemOpen
  \bibfield  {author} {\bibinfo {author} {\bibfnamefont {S.~R.~K.}\
  \bibnamefont {Rodriguez}}, \bibinfo {author} {\bibfnamefont {W.}~\bibnamefont
  {Casteels}}, \bibinfo {author} {\bibfnamefont {F.}~\bibnamefont {Storme}},
  \bibinfo {author} {\bibfnamefont {N.}~\bibnamefont {Carlon~Zambon}}, \bibinfo
  {author} {\bibfnamefont {I.}~\bibnamefont {Sagnes}}, \bibinfo {author}
  {\bibfnamefont {L.}~\bibnamefont {Le~Gratiet}}, \bibinfo {author}
  {\bibfnamefont {E.}~\bibnamefont {Galopin}}, \bibinfo {author} {\bibfnamefont
  {A.}~\bibnamefont {Lema\^{\i}tre}}, \bibinfo {author} {\bibfnamefont
  {A.}~\bibnamefont {Amo}}, \bibinfo {author} {\bibfnamefont {C.}~\bibnamefont
  {Ciuti}},\ and\ \bibinfo {author} {\bibfnamefont {J.}~\bibnamefont {Bloch}},\
  }\bibfield  {title} {\bibinfo {title} {Probing a dissipative phase transition
  via dynamical optical hysteresis},\ }\href
  {https://doi.org/10.1103/PhysRevLett.118.247402} {\bibfield  {journal}
  {\bibinfo  {journal} {Phys. Rev. Lett.}\ }\textbf {\bibinfo {volume} {118}},\
  \bibinfo {pages} {247402} (\bibinfo {year} {2017})}\BibitemShut {NoStop}%
\bibitem [{\citenamefont {Laussy}\ \emph {et~al.}(2010)\citenamefont {Laussy},
  \citenamefont {Kavokin},\ and\ \citenamefont {Shelykh}}]{Laussy10}%
  \BibitemOpen
  \bibfield  {author} {\bibinfo {author} {\bibfnamefont {F.~P.}\ \bibnamefont
  {Laussy}}, \bibinfo {author} {\bibfnamefont {A.~V.}\ \bibnamefont
  {Kavokin}},\ and\ \bibinfo {author} {\bibfnamefont {I.~A.}\ \bibnamefont
  {Shelykh}},\ }\bibfield  {title} {\bibinfo {title} {{E}xciton-{P}olariton
  {M}ediated {S}uperconductivity},\ }\href
  {https://doi.org/10.1103/PhysRevLett.104.106402} {\bibfield  {journal}
  {\bibinfo  {journal} {Phys. Rev. Lett.}\ }\textbf {\bibinfo {volume} {104}},\
  \bibinfo {pages} {106402} (\bibinfo {year} {2010})}\BibitemShut {NoStop}%
\bibitem [{\citenamefont {Orgiu}\ \emph {et~al.}(2015)\citenamefont {Orgiu},
  \citenamefont {George}, \citenamefont {Hutchison}, \citenamefont {Devaux},
  \citenamefont {Dayen}, \citenamefont {Doudin}, \citenamefont {Stellacci},
  \citenamefont {Genet}, \citenamefont {Schachenmayer}, \citenamefont {Genes},
  \citenamefont {Pupillo}, \citenamefont {Samor\'{\i}},\ and\ \citenamefont
  {Ebbesen}}]{Ebbesen15}%
  \BibitemOpen
  \bibfield  {author} {\bibinfo {author} {\bibfnamefont {E.}~\bibnamefont
  {Orgiu}}, \bibinfo {author} {\bibfnamefont {J.}~\bibnamefont {George}},
  \bibinfo {author} {\bibfnamefont {J.}~\bibnamefont {Hutchison}}, \bibinfo
  {author} {\bibfnamefont {E.}~\bibnamefont {Devaux}}, \bibinfo {author}
  {\bibfnamefont {J.}~\bibnamefont {Dayen}}, \bibinfo {author} {\bibfnamefont
  {B.}~\bibnamefont {Doudin}}, \bibinfo {author} {\bibfnamefont
  {F.}~\bibnamefont {Stellacci}}, \bibinfo {author} {\bibfnamefont
  {C.}~\bibnamefont {Genet}}, \bibinfo {author} {\bibfnamefont
  {J.}~\bibnamefont {Schachenmayer}}, \bibinfo {author} {\bibfnamefont
  {C.}~\bibnamefont {Genes}}, \bibinfo {author} {\bibfnamefont
  {G.}~\bibnamefont {Pupillo}}, \bibinfo {author} {\bibfnamefont
  {P.}~\bibnamefont {Samor\'{\i}}},\ and\ \bibinfo {author} {\bibfnamefont
  {T.}~\bibnamefont {Ebbesen}},\ }\bibfield  {title} {\bibinfo {title}
  {{C}onductivity in organic semiconductors hybridized with the vacuum field},\
  }\href {https://www.nature.com/articles/nmat4392} {\bibfield  {journal}
  {\bibinfo  {journal} {Nat. Mater.}\ }\textbf {\bibinfo {volume} {14}},\
  \bibinfo {pages} {1123} (\bibinfo {year} {2015})}\BibitemShut {NoStop}%
\bibitem [{\citenamefont {Coles}\ \emph {et~al.}(2014)\citenamefont {Coles},
  \citenamefont {Yang}, \citenamefont {Wang}, \citenamefont {Grant},
  \citenamefont {Taylor}, \citenamefont {Saikin}, \citenamefont {Aspuru-Guzik},
  \citenamefont {Lidzey}, \citenamefont {Tang},\ and\ \citenamefont
  {Smith}}]{Coles14}%
  \BibitemOpen
  \bibfield  {author} {\bibinfo {author} {\bibfnamefont {D.~M.}\ \bibnamefont
  {Coles}}, \bibinfo {author} {\bibfnamefont {Y.}~\bibnamefont {Yang}},
  \bibinfo {author} {\bibfnamefont {Y.}~\bibnamefont {Wang}}, \bibinfo {author}
  {\bibfnamefont {R.~T.}\ \bibnamefont {Grant}}, \bibinfo {author}
  {\bibfnamefont {R.~A.}\ \bibnamefont {Taylor}}, \bibinfo {author}
  {\bibfnamefont {S.~K.}\ \bibnamefont {Saikin}}, \bibinfo {author}
  {\bibfnamefont {A.}~\bibnamefont {Aspuru-Guzik}}, \bibinfo {author}
  {\bibfnamefont {D.~G.}\ \bibnamefont {Lidzey}}, \bibinfo {author}
  {\bibfnamefont {J.~K.-H.}\ \bibnamefont {Tang}},\ and\ \bibinfo {author}
  {\bibfnamefont {J.~M.}\ \bibnamefont {Smith}},\ }\bibfield  {title} {\bibinfo
  {title} {{S}trong coupling between chlorosomes of photosynthetic bacteria and
  a confined optical cavity mode},\ }\href
  {https://www.nature.com/articles/ncomms6561} {\bibfield  {journal} {\bibinfo
  {journal} {Nat. Commun.}\ }\textbf {\bibinfo {volume} {5}},\ \bibinfo {pages}
  {1} (\bibinfo {year} {2014})}\BibitemShut {NoStop}%
\bibitem [{\citenamefont {Su}\ \emph {et~al.}(2017)\citenamefont {Su},
  \citenamefont {Diederichs}, \citenamefont {Wang}, \citenamefont {Liew},
  \citenamefont {Zhao}, \citenamefont {Liu}, \citenamefont {Xu}, \citenamefont
  {Chen},\ and\ \citenamefont {Xiong}}]{Su17}%
  \BibitemOpen
  \bibfield  {author} {\bibinfo {author} {\bibfnamefont {R.}~\bibnamefont
  {Su}}, \bibinfo {author} {\bibfnamefont {C.}~\bibnamefont {Diederichs}},
  \bibinfo {author} {\bibfnamefont {J.}~\bibnamefont {Wang}}, \bibinfo {author}
  {\bibfnamefont {T.~C.~H.}\ \bibnamefont {Liew}}, \bibinfo {author}
  {\bibfnamefont {J.}~\bibnamefont {Zhao}}, \bibinfo {author} {\bibfnamefont
  {S.}~\bibnamefont {Liu}}, \bibinfo {author} {\bibfnamefont {W.}~\bibnamefont
  {Xu}}, \bibinfo {author} {\bibfnamefont {Z.}~\bibnamefont {Chen}},\ and\
  \bibinfo {author} {\bibfnamefont {Q.}~\bibnamefont {Xiong}},\ }\bibfield
  {title} {\bibinfo {title} {{R}oom-{T}emperature {P}olariton {L}asing in
  {A}ll-{I}norganic {P}erovskite {N}anoplatelets},\ }\href
  {https://doi.org/10.1021/acs.nanolett.7b01956} {\bibfield  {journal}
  {\bibinfo  {journal} {Nano Lett.}\ }\textbf {\bibinfo {volume} {17}},\
  \bibinfo {pages} {3982} (\bibinfo {year} {2017})}\BibitemShut {NoStop}%
\bibitem [{\citenamefont {Ramezani}\ \emph {et~al.}(2017)\citenamefont
  {Ramezani}, \citenamefont {Halpin}, \citenamefont
  {Fern\'{a}ndez-Dom\'{i}nguez}, \citenamefont {Feist}, \citenamefont
  {Rodriguez}, \citenamefont {Garcia-Vidal},\ and\ \citenamefont
  {Rivas}}]{Ramezani17}%
  \BibitemOpen
  \bibfield  {author} {\bibinfo {author} {\bibfnamefont {M.}~\bibnamefont
  {Ramezani}}, \bibinfo {author} {\bibfnamefont {A.}~\bibnamefont {Halpin}},
  \bibinfo {author} {\bibfnamefont {A.~I.}\ \bibnamefont
  {Fern\'{a}ndez-Dom\'{i}nguez}}, \bibinfo {author} {\bibfnamefont
  {J.}~\bibnamefont {Feist}}, \bibinfo {author} {\bibfnamefont {S.~R.-K.}\
  \bibnamefont {Rodriguez}}, \bibinfo {author} {\bibfnamefont {F.~J.}\
  \bibnamefont {Garcia-Vidal}},\ and\ \bibinfo {author} {\bibfnamefont {J.~G.}\
  \bibnamefont {Rivas}},\ }\bibfield  {title} {\bibinfo {title}
  {Plasmon-exciton-polariton lasing},\ }\href
  {https://doi.org/10.1364/OPTICA.4.000031} {\bibfield  {journal} {\bibinfo
  {journal} {Optica}\ }\textbf {\bibinfo {volume} {4}},\ \bibinfo {pages} {31}
  (\bibinfo {year} {2017})}\BibitemShut {NoStop}%
\bibitem [{\citenamefont {Delteil}\ \emph {et~al.}(2019)\citenamefont
  {Delteil}, \citenamefont {Fink}, \citenamefont {Schade}, \citenamefont
  {H{\"o}fling}, \citenamefont {Schneider},\ and\ \citenamefont
  {{\.I}mamo{\u{g}}lu}}]{Delteil19}%
  \BibitemOpen
  \bibfield  {author} {\bibinfo {author} {\bibfnamefont {A.}~\bibnamefont
  {Delteil}}, \bibinfo {author} {\bibfnamefont {T.}~\bibnamefont {Fink}},
  \bibinfo {author} {\bibfnamefont {A.}~\bibnamefont {Schade}}, \bibinfo
  {author} {\bibfnamefont {S.}~\bibnamefont {H{\"o}fling}}, \bibinfo {author}
  {\bibfnamefont {C.}~\bibnamefont {Schneider}},\ and\ \bibinfo {author}
  {\bibfnamefont {A.}~\bibnamefont {{\.I}mamo{\u{g}}lu}},\ }\bibfield  {title}
  {\bibinfo {title} {{T}owards polariton blockade of confined
  exciton--polaritons},\ }\href
  {https://www.nature.com/articles/s41563-019-0282-y} {\bibfield  {journal}
  {\bibinfo  {journal} {Nat. Mater}\ }\textbf {\bibinfo {volume} {18}},\
  \bibinfo {pages} {219} (\bibinfo {year} {2019})}\BibitemShut {NoStop}%
\bibitem [{\citenamefont {Mu{\~n}oz-Matutano}\ \emph
  {et~al.}(2019)\citenamefont {Mu{\~n}oz-Matutano}, \citenamefont {Wood},
  \citenamefont {Johnsson}, \citenamefont {Vidal}, \citenamefont {Baragiola},
  \citenamefont {Reinhard}, \citenamefont {Lema{\^\i}tre}, \citenamefont
  {Bloch}, \citenamefont {Amo}, \citenamefont {Nogues}, \citenamefont {Besga},
  \citenamefont {Richard},\ and\ \citenamefont {Volz}}]{Volz19}%
  \BibitemOpen
  \bibfield  {author} {\bibinfo {author} {\bibfnamefont {G.}~\bibnamefont
  {Mu{\~n}oz-Matutano}}, \bibinfo {author} {\bibfnamefont {A.}~\bibnamefont
  {Wood}}, \bibinfo {author} {\bibfnamefont {M.}~\bibnamefont {Johnsson}},
  \bibinfo {author} {\bibfnamefont {X.}~\bibnamefont {Vidal}}, \bibinfo
  {author} {\bibfnamefont {B.~Q.}\ \bibnamefont {Baragiola}}, \bibinfo {author}
  {\bibfnamefont {A.}~\bibnamefont {Reinhard}}, \bibinfo {author}
  {\bibfnamefont {A.}~\bibnamefont {Lema{\^\i}tre}}, \bibinfo {author}
  {\bibfnamefont {J.}~\bibnamefont {Bloch}}, \bibinfo {author} {\bibfnamefont
  {A.}~\bibnamefont {Amo}}, \bibinfo {author} {\bibfnamefont {G.}~\bibnamefont
  {Nogues}}, \bibinfo {author} {\bibfnamefont {B.}~\bibnamefont {Besga}},
  \bibinfo {author} {\bibfnamefont {M.}~\bibnamefont {Richard}},\ and\ \bibinfo
  {author} {\bibfnamefont {T.}~\bibnamefont {Volz}},\ }\bibfield  {title}
  {\bibinfo {title} {{E}mergence of quantum correlations from interacting
  fibre-cavity polaritons},\ }\href
  {https://www.nature.com/articles/s41563-019-0281-z} {\bibfield  {journal}
  {\bibinfo  {journal} {Nat. Mater.}\ }\textbf {\bibinfo {volume} {18}},\
  \bibinfo {pages} {213} (\bibinfo {year} {2019})}\BibitemShut {NoStop}%
\bibitem [{\citenamefont {Savona}\ \emph {et~al.}(1995)\citenamefont {Savona},
  \citenamefont {Andreani}, \citenamefont {Schwendimann},\ and\ \citenamefont
  {Quattropani}}]{Savona95}%
  \BibitemOpen
  \bibfield  {author} {\bibinfo {author} {\bibfnamefont {V.}~\bibnamefont
  {Savona}}, \bibinfo {author} {\bibfnamefont {L.~C.}\ \bibnamefont
  {Andreani}}, \bibinfo {author} {\bibfnamefont {P.}~\bibnamefont
  {Schwendimann}},\ and\ \bibinfo {author} {\bibfnamefont {A.}~\bibnamefont
  {Quattropani}},\ }\bibfield  {title} {\bibinfo {title} {{Q}uantum well
  excitons in semiconductor microcavities: {U}nified treatment of weak and
  strong coupling regimes},\ }\href
  {https://doi.org/https://doi.org/10.1016/0038-1098(94)00865-5} {\bibfield
  {journal} {\bibinfo  {journal} {Solid State Commun.}\ }\textbf {\bibinfo
  {volume} {93}},\ \bibinfo {pages} {733} (\bibinfo {year} {1995})}\BibitemShut
  {NoStop}%
\bibitem [{\citenamefont {Schwartz}\ \emph {et~al.}(2011)\citenamefont
  {Schwartz}, \citenamefont {Hutchison}, \citenamefont {Genet},\ and\
  \citenamefont {Ebbesen}}]{Ebbesen11}%
  \BibitemOpen
  \bibfield  {author} {\bibinfo {author} {\bibfnamefont {T.}~\bibnamefont
  {Schwartz}}, \bibinfo {author} {\bibfnamefont {J.~A.}\ \bibnamefont
  {Hutchison}}, \bibinfo {author} {\bibfnamefont {C.}~\bibnamefont {Genet}},\
  and\ \bibinfo {author} {\bibfnamefont {T.~W.}\ \bibnamefont {Ebbesen}},\
  }\bibfield  {title} {\bibinfo {title} {{R}eversible {S}witching of
  {U}ltrastrong {L}ight-{M}olecule {C}oupling},\ }\href
  {https://doi.org/10.1103/PhysRevLett.106.196405} {\bibfield  {journal}
  {\bibinfo  {journal} {Phys. Rev. Lett.}\ }\textbf {\bibinfo {volume} {106}},\
  \bibinfo {pages} {196405} (\bibinfo {year} {2011})}\BibitemShut {NoStop}%
\bibitem [{\citenamefont {Melnikau}\ \emph {et~al.}(2016)\citenamefont
  {Melnikau}, \citenamefont {Esteban}, \citenamefont {Savateeva}, \citenamefont
  {Sánchez-Iglesias}, \citenamefont {Grzelczak}, \citenamefont {Schmidt},
  \citenamefont {Liz-Marzán}, \citenamefont {Aizpurua},\ and\ \citenamefont
  {Rakovich}}]{Rakovich}%
  \BibitemOpen
  \bibfield  {author} {\bibinfo {author} {\bibfnamefont {D.}~\bibnamefont
  {Melnikau}}, \bibinfo {author} {\bibfnamefont {R.}~\bibnamefont {Esteban}},
  \bibinfo {author} {\bibfnamefont {D.}~\bibnamefont {Savateeva}}, \bibinfo
  {author} {\bibfnamefont {A.}~\bibnamefont {Sánchez-Iglesias}}, \bibinfo
  {author} {\bibfnamefont {M.}~\bibnamefont {Grzelczak}}, \bibinfo {author}
  {\bibfnamefont {M.~K.}\ \bibnamefont {Schmidt}}, \bibinfo {author}
  {\bibfnamefont {L.~M.}\ \bibnamefont {Liz-Marzán}}, \bibinfo {author}
  {\bibfnamefont {J.}~\bibnamefont {Aizpurua}},\ and\ \bibinfo {author}
  {\bibfnamefont {Y.~P.}\ \bibnamefont {Rakovich}},\ }\bibfield  {title}
  {\bibinfo {title} {Rabi splitting in photoluminescence spectra of hybrid
  systems of gold nanorods and {J}-aggregates},\ }\href
  {https://doi.org/10.1021/acs.jpclett.5b02512} {\bibfield  {journal} {\bibinfo
   {journal} {J. Phys. Chem. Lett.}\ }\textbf {\bibinfo {volume} {7}},\
  \bibinfo {pages} {354} (\bibinfo {year} {2016})}\BibitemShut {NoStop}%
\bibitem [{\citenamefont {Rodriguez}\ \emph {et~al.}(2012)\citenamefont
  {Rodriguez}, \citenamefont {Murai}, \citenamefont {Verschuuren},\ and\
  \citenamefont {Rivas}}]{Rodriguez12}%
  \BibitemOpen
  \bibfield  {author} {\bibinfo {author} {\bibfnamefont {S.~R.~K.}\
  \bibnamefont {Rodriguez}}, \bibinfo {author} {\bibfnamefont {S.}~\bibnamefont
  {Murai}}, \bibinfo {author} {\bibfnamefont {M.~A.}\ \bibnamefont
  {Verschuuren}},\ and\ \bibinfo {author} {\bibfnamefont {J.~G.}\ \bibnamefont
  {Rivas}},\ }\bibfield  {title} {\bibinfo {title} {Light-emitting
  waveguide-plasmon polaritons},\ }\href
  {https://doi.org/10.1103/PhysRevLett.109.166803} {\bibfield  {journal}
  {\bibinfo  {journal} {Phys. Rev. Lett.}\ }\textbf {\bibinfo {volume} {109}},\
  \bibinfo {pages} {166803} (\bibinfo {year} {2012})}\BibitemShut {NoStop}%
\bibitem [{\citenamefont {Du}\ \emph {et~al.}(2018)\citenamefont {Du},
  \citenamefont {Zhang}, \citenamefont {Shi}, \citenamefont {Chen},
  \citenamefont {Wu}, \citenamefont {Mi}, \citenamefont {Liu}, \citenamefont
  {Li}, \citenamefont {Sui}, \citenamefont {Wang}, \citenamefont {Qiu},
  \citenamefont {Wu}, \citenamefont {Xiao}, \citenamefont {Zhang},\ and\
  \citenamefont {Liu}}]{Liu18}%
  \BibitemOpen
  \bibfield  {author} {\bibinfo {author} {\bibfnamefont {W.}~\bibnamefont
  {Du}}, \bibinfo {author} {\bibfnamefont {S.}~\bibnamefont {Zhang}}, \bibinfo
  {author} {\bibfnamefont {J.}~\bibnamefont {Shi}}, \bibinfo {author}
  {\bibfnamefont {J.}~\bibnamefont {Chen}}, \bibinfo {author} {\bibfnamefont
  {Z.}~\bibnamefont {Wu}}, \bibinfo {author} {\bibfnamefont {Y.}~\bibnamefont
  {Mi}}, \bibinfo {author} {\bibfnamefont {Z.}~\bibnamefont {Liu}}, \bibinfo
  {author} {\bibfnamefont {Y.}~\bibnamefont {Li}}, \bibinfo {author}
  {\bibfnamefont {X.}~\bibnamefont {Sui}}, \bibinfo {author} {\bibfnamefont
  {R.}~\bibnamefont {Wang}}, \bibinfo {author} {\bibfnamefont {X.}~\bibnamefont
  {Qiu}}, \bibinfo {author} {\bibfnamefont {T.}~\bibnamefont {Wu}}, \bibinfo
  {author} {\bibfnamefont {Y.}~\bibnamefont {Xiao}}, \bibinfo {author}
  {\bibfnamefont {Q.}~\bibnamefont {Zhang}},\ and\ \bibinfo {author}
  {\bibfnamefont {X.}~\bibnamefont {Liu}},\ }\bibfield  {title} {\bibinfo
  {title} {Strong exciton–photon coupling and lasing behavior in
  all-inorganic {C}s{P}b{B}r3 micro/nanowire {F}abry-{P}érot cavity},\ }\href
  {https://doi.org/10.1021/acsphotonics.7b01593} {\bibfield  {journal}
  {\bibinfo  {journal} {ACS Photonics}\ }\textbf {\bibinfo {volume} {5}},\
  \bibinfo {pages} {2051} (\bibinfo {year} {2018})}\BibitemShut {NoStop}%
\bibitem [{\citenamefont {Su}\ \emph {et~al.}(2020)\citenamefont {Su},
  \citenamefont {Ghosh}, \citenamefont {Wang}, \citenamefont {Liu},
  \citenamefont {Diederichs}, \citenamefont {Liew},\ and\ \citenamefont
  {Xiong}}]{Xiong20}%
  \BibitemOpen
  \bibfield  {author} {\bibinfo {author} {\bibfnamefont {R.}~\bibnamefont
  {Su}}, \bibinfo {author} {\bibfnamefont {S.}~\bibnamefont {Ghosh}}, \bibinfo
  {author} {\bibfnamefont {J.}~\bibnamefont {Wang}}, \bibinfo {author}
  {\bibfnamefont {S.}~\bibnamefont {Liu}}, \bibinfo {author} {\bibfnamefont
  {C.}~\bibnamefont {Diederichs}}, \bibinfo {author} {\bibfnamefont {T.~C.}\
  \bibnamefont {Liew}},\ and\ \bibinfo {author} {\bibfnamefont
  {Q.}~\bibnamefont {Xiong}},\ }\bibfield  {title} {\bibinfo {title}
  {{O}bservation of exciton polariton condensation in a perovskite lattice at
  room temperature},\ }\href
  {https://www.nature.com/articles/s41567-019-0764-5} {\bibfield  {journal}
  {\bibinfo  {journal} {Nat. Phys.}\ }\textbf {\bibinfo {volume} {16}},\
  \bibinfo {pages} {301} (\bibinfo {year} {2020})}\BibitemShut {NoStop}%
\bibitem [{\citenamefont {Zhang}\ \emph {et~al.}(2016)\citenamefont {Zhang},
  \citenamefont {Su}, \citenamefont {Liu}, \citenamefont {Xing}, \citenamefont
  {Sum},\ and\ \citenamefont {Xiong}}]{Xiong16}%
  \BibitemOpen
  \bibfield  {author} {\bibinfo {author} {\bibfnamefont {Q.}~\bibnamefont
  {Zhang}}, \bibinfo {author} {\bibfnamefont {R.}~\bibnamefont {Su}}, \bibinfo
  {author} {\bibfnamefont {X.}~\bibnamefont {Liu}}, \bibinfo {author}
  {\bibfnamefont {J.}~\bibnamefont {Xing}}, \bibinfo {author} {\bibfnamefont
  {T.~C.}\ \bibnamefont {Sum}},\ and\ \bibinfo {author} {\bibfnamefont
  {Q.}~\bibnamefont {Xiong}},\ }\bibfield  {title} {\bibinfo {title}
  {{H}igh-{Q}uality {W}hispering-{G}allery-{M}ode {L}asing from {C}esium {L}ead
  {H}alide {P}erovskite {N}anoplatelets},\ }\href
  {https://doi.org/10.1002/adfm.201601690} {\bibfield  {journal} {\bibinfo
  {journal} {Adv. Funct. Mater.}\ }\textbf {\bibinfo {volume} {26}},\ \bibinfo
  {pages} {6238} (\bibinfo {year} {2016})}\BibitemShut {NoStop}%
\bibitem [{\citenamefont {Su}\ \emph {et~al.}(2018)\citenamefont {Su},
  \citenamefont {Wang}, \citenamefont {Zhao}, \citenamefont {Xing},
  \citenamefont {Zhao}, \citenamefont {Diederichs}, \citenamefont {Liew},\ and\
  \citenamefont {Xiong}}]{Xiong18}%
  \BibitemOpen
  \bibfield  {author} {\bibinfo {author} {\bibfnamefont {R.}~\bibnamefont
  {Su}}, \bibinfo {author} {\bibfnamefont {J.}~\bibnamefont {Wang}}, \bibinfo
  {author} {\bibfnamefont {J.}~\bibnamefont {Zhao}}, \bibinfo {author}
  {\bibfnamefont {J.}~\bibnamefont {Xing}}, \bibinfo {author} {\bibfnamefont
  {W.}~\bibnamefont {Zhao}}, \bibinfo {author} {\bibfnamefont {C.}~\bibnamefont
  {Diederichs}}, \bibinfo {author} {\bibfnamefont {T.~C.}\ \bibnamefont
  {Liew}},\ and\ \bibinfo {author} {\bibfnamefont {Q.}~\bibnamefont {Xiong}},\
  }\bibfield  {title} {\bibinfo {title} {Room temperature long-range coherent
  exciton polariton condensate flow in lead halide perovskites},\ }\href
  {https://advances.sciencemag.org/content/4/10/eaau0244} {\bibfield  {journal}
  {\bibinfo  {journal} {Sci. Adv.}\ }\textbf {\bibinfo {volume} {4}},\ \bibinfo
  {pages} {eaau0244} (\bibinfo {year} {2018})}\BibitemShut {NoStop}%
\end{thebibliography}

%

\end{document}